\documentclass[a4paper, oneside, twocolumn, notitlepage, 10pt]{extarticle_ecoc2015}
\usepackage{ecoc2015}
\usepackage{tikz}
\usepackage{pgfplots}
\pgfplotsset{compat=newest} 
\pgfplotsset{plot coordinates/math parser=false}
\usepackage{graphicx}
\usepackage{booktabs}

\begin{document}
\newlength{\picwidth}
\newlength{\picheight}


\title{
Demonstration of Fully Nonlinear Spectrum Modulated System in the Highly Nonlinear Optical Transmission Regime 
}
%


\author{
    Vahid Aref, Son T. Le, and Henning Buelow
}

\maketitle                  


\begin{strip}
 \begin{author_descr}

   Nokia Bell Labs, Lorenzstr. 10, 70435 Stuttgart, Germany,
  \uline{firstname.lastname@nokia-bell-labs.com}

 \end{author_descr}
\end{strip}

\setstretch{1.04}


\begin{strip}
  \begin{ecoc_abstract}
We report a 3 dB increase in the nonlinear threshold of a 64 $\times$ 0.5Gbaud 16-QAM continuous-nonlinear-spectrum modulated signal by nonlinear multiplexing with QPSK modulated multi-solitons, showing the first ever fully nonlinear-spectrum modulated system in the highly nonlinear regime. 
  \end{ecoc_abstract}
\end{strip}

\section{Introduction}
Recent advances 
in coherent technology and applying digital signal processing techniques, originally developed for
classical linear channels,
enable us to achieve the spectral efficiency (SE) close to the ``presumed''
upper-bound of AWGN capacity for low power signals in which the channel is fairly linear.  
However, the achievable SE
decreases
for launch powers beyond some ``nonlinear threshold'' as the channel becomes very nonlinear. 
The nonlinear threshold depends on modulation formats, pulse-shaping, and can be improved by more complex equalization techniques such as digital back propagation.


Nonlinear Fourier transform (NFT) characterizes the complex signal propagation in optical fiber
by some simple evolution rules. It maps each pulse in a nonlinear spectrum, partitioned to continuous and discrete part. Continuous spectrum represents the ``dispersive'' part of the pulse
while 
the discrete spectrum represents 
the ``solitonic'' part, 
nonlinearly multiplexed to the dispersive part. Solitonic part appears when the pulse has a large power.

As a pulse propagates in a lossless-noiseless fiber, 
the modes of nonlinear spectrum are independently evolved. 
This fact proposes to modulate data over these 
modes.
Accordingly,
the modulation of OFDM symbols in the continuous spectrum is demonstrated 
in \cite{le2015nonlinear} showing the potential performance gain
over the classical OFDM symbols. However, the performance still degrades by increasing
power beyond some nonlinear threshold. Moreover,
modulation of pure solitonic pulses (without continuous part) is
 demonstrated in 
\cite{aref2015experimental}, \cite{dong2015nonlinear}, \cite{Buelow20167eigenvalues}.
However, the achieved SE is so far very low, 
and multi-solitons are observed susceptible 
to imperfect transmission conditions.

In this paper, we demonstrate for the first time the modulation over both discrete 
and continuous parts by developing a low-complexity inverse NFT (INFT) algorithm. On the continuous spectrum,
we modulate $64\times 0.5$ Gbaud OFDM signals with $16$-QAM format 
and on the discrete spectrum, we consider a hybrid modulation of eigenvalues and spectral amplitudes: First two eigenvalues are chosen out of a set of eigenvalues and then, the spectral amplitude of each chosen eigenvalue is QPSK modulated. These two QPSK modulated solitons, carrying additional few information bits, are multiplexed with the signal of continuous part. 

We demonstrate the transmission over $18\times 81.3$ km in a SMF fiber loop with EDFAs. 
The signal of each spectrum are detected independently in nonlinear spectrum.
We show that the nonlinear threshold will be increased proportional to the additional power in
discrete spectrum. However,
the performance of continuous spectrum is slightly degraded
by multiplexing multi-solitons. 
This can be because of larger pick-to-average power ratio, and more importantly, the weak correlation between two spectra caused by the noise of amplifiers and by the lumped amplification.  

%
\begin{figure}[tb!]
   \centering
   \input{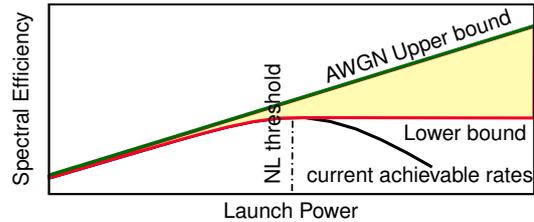}
    \caption{\label{fig:se_region} The spectral efficiency of a typical 
    coherent optical system. 
    }
    
\end{figure}

\section{Modulation of Discrete and Continuous Spectrum}

\begin{figure}[tb!]
   \centering
   \input{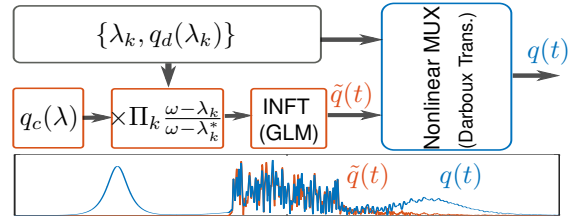}
        \vspace{2pt}
    \caption{\label{fig:mux}Multiplexing discrete and continuous spectrum}
    
\end{figure}

The nonlinear spectrum of a pulse $q(t)$ has two parts: The continuous part, denoted by $q_c(\lambda)$,
is the spectral amplitudes for real valued frequencies $\lambda\in\mathbb{R}$. The discrete part
contains a finite set of complex eigenvalues $\{\lambda_k\}\subset\mathbb{C}^+$ and the
corresponding spectral amplitudes $q_d(\lambda_k)$. The spectral amplitudes are defined based on
Zakharov-Shabat system. We refer the reader to
\cite{yousefi2014information} for detailed definitions and properties. We modulate information bits
over both spectrum as follows:
\vspace{4pt}
\\
\emph{\textit{A. Continuous Spectrum:}} We modulate $256$ bits over 
$N=64$ overlapping orthogonal subchannels in the continuous spectrum in a similar way as the conventional OFDM signal is designed in the linear Fourier spectrum. For each symbol, the continuous spectrum was defined as:
\begin{equation}\label{eq:qc}\textstyle
q_c(\omega) = A \sum_{k=-N/2}^{N/2} C_k{\rm sinc}(\frac{T_c}{\pi}\omega +k), \omega\in\mathbb{R},
\end{equation}
where $C_k$ is drawn from $16$-QAM constellation, $A$ is the power control parameter, $T_c = 2$ ns is the useful block duration defining the baud-rate of 0.5Gbaud. 
To avoid ISI during propagation, we consider two guard intervals of $4$ ns, resulting the total symbol duration $T_s=10$ ns and the raw data rate of $25.6$ Gb/s modulated in continuous part. 
\vspace{4pt}
\\
\emph{\textit{B. Discrete Spectrum:}} 
We modulate the position of two purely imaginary eigenvalues $\lambda_1$ and $\lambda_2$ in the upper complex plane as well as the phase of their spectral amplitudes. Each circle in Fig.~\ref{fig:dis_spec} (a) illustrates a pair of the 12 possible
$(\lambda_1,\lambda_2)$
which are used for modulation. In addition, the phase of each spectral amplitude, 
$\sphericalangle q_d(\lambda_k)$, is independently modulated with a QPSK constellation, as shown in Fig.~\ref{fig:dis_spec} (b). 
Determining the temporal position of each ``sech'' shape in the symbol (see Fig.~\ref{fig:mux}),  $|q_d(\lambda_k)|$ are chosen 
such that a ``sech'' shape occurs every almost 5 ns in the block of symbols. We denote this modulation 
by $D2$. 

We also consider another modulation scheme, denoted here by $D1$, with the different 
eigenvalue set of $\{1j,1.5j\}$. One bit is thus modulated by
two possible pairs of $(\lambda_1,\lambda_2)$. The spectral amplitudes are modulated similar to $D2$. 
The average power of discrete spectrum for each of modulations is,
\begin{equation}\label{eq:d}
\begin{aligned}\textstyle
P_{\rm d}&=\textstyle\frac{|\beta_2|}{\gamma T_0^2}\approx\textstyle 0.016 \text{ mW, for } D1\\
P_{\rm d}&=2\textstyle\frac{|\beta_2|}{\gamma T_0^2}\approx 0.032 \text{ mW, for } D2
\end{aligned}
\end{equation}
for $\beta_2=-21.3$ $\frac{\text{ps}^2}{km}$, $\gamma=1.3 \frac{\text{W}^{-1}}{\text{km}}$,
and scaling factor $T_0=1$ ns. 
%
%

Moreover,  5 bits are modulated per symbol in $D1$ and $4+\log_2(12)$ bits in $D2$, resulting the rate of 0.5Gb/s and 
0.76 Gb/s, respectively.
\vspace{4pt}
\\
\emph{\textit{C. Multiplexing Spectrum:}} 
To multiplex both spectrum, denoted by $(C\boxplus D)$, we extend the Darboux transformation used to generate a multi-soliton pulse. 
It initially
starts with $q(t)=0$ and updates $q(t)$ by adding eigenvalue one-by-one.
The algorithm and how nonlinear spectrum is changed by updating $q(t)$ are described in \citep{aref2016control}. We 
apply the Darboux transformation with initial pulse $q(t)=\tilde{q}(t)\neq 0$. 
To add the eigenvalue $\lambda_k$,
we need a solution of Zakharov-Shabat system at this spectral parameter $\lambda_k$. 
For the sake of space, we skip here how one can find a suitable solution. Having this solution,
we can update $q(t)$  to a pulse having eigenvalue $\lambda_k$. 
According to \cite[Theorem 1]{aref2016control}, the continuous spectrum of the pulse after adding $n$ eigenvalues $\{\lambda_1,\dots,\lambda_n\}$ is,
\begin{equation}\label{eq:evolution}
q_c(\omega)= \tilde{q}_c(\omega)\textstyle\prod_{k=1}^n \textstyle\frac{\omega-\lambda_k^*}{\omega-\lambda_k},
\end{equation} 
where $\tilde{q}_c(\omega)$ denotes the continuous
spectrum of the initial pulse $\tilde{q}(t)$. To have the given $q_c(\omega)$ in Eq. \ref{eq:qc}, we first 
apply the inverse of evolution Eq. \ref{eq:evolution}. For taking inverse NFT, we then use the Gelfand-Levitan-Marchenko (GLM) algorithm~\cite{le2015nonlinear} to find $\tilde{q}(t)$ and finally
apply the Darboux transformation. The diagram of the algorithm including a pulse of $(C\boxplus D2)$ are shown in Fig.~\ref{fig:mux}.  

\begin{figure}[tb!]
   \centering
   \input{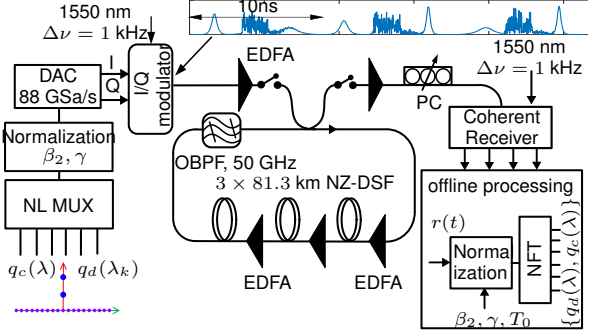}
        \vspace{1pt}
        
    \caption{\label{fig:setup}Experimental setup with offline NFT-based detection}
\end{figure}

\section{Experimental results}

The experimental setup with a re-circulating loop is illustrated in
Fig.~\ref{fig:setup}. The time domain signal $q(t)$ was generated 
from multiplexing symbol by symbol random streams of continuous and discrete spectrum having the total power, $P= P_{\rm d} + P_{\rm c}$
where $P_{\rm d}$ is fixed as Eq. \ref{eq:d}, and $P_{\rm c}$ is controlled by parameter $A$ in Eq. \ref{eq:qc}.
After resampling to 88 GSa/s, $q(t)$ is normalized according to the path averaged model of links with lumped amplifiers.

The fiber loop consists of $3\times 81.3$ km spans of standard single mode fiber and EDFAs. Both the transmitter laser and local oscillator were from a single fiber laser source with $1$ kHz linewidth. At the receiver, after coherent detection, digital sampling at 80 GSa/s, timing synchronization and frequency offset compensation; the received signal was normalized to the initial power $P$. Next, for each 10 ns symbol, NFT was applied separately on each spectrum as follows:
\vspace{4pt}
\\
\emph{\textit{A. Continuous Spectrum:}}  The detection of $64\times 0.5$GBaud $16$-QAM OFDM symbols were carried out according to \cite{le2015nonlinear} in which
single-tap phase-shift removal operation was performed to remove the interplay of dispersion and nonlinearity followed then by channel equalization and phase noise estimation.
\begin{figure}[tb!]
\setlength{\picwidth}{0.50\textwidth}
\setlength{\picheight}{0.45\textwidth}
  \centering
\definecolor{mycolor1}{rgb}{0.00000,0.44700,0.74100}%
\definecolor{mycolor2}{rgb}{0.85000,0.32500,0.09800}%
\definecolor{mycolor3}{rgb}{0.92900,0.69400,0.12500}%
\definecolor{mycolor4}{rgb}{0.49400,0.18400,0.55600}%
\definecolor{mycolor5}{rgb}{0.46600,0.67400,0.18800}%
\definecolor{mycolor6}{rgb}{0.30100,0.74500,0.93300}%
\definecolor{mycolor7}{rgb}{0.63500,0.07800,0.18400}%
\begin{tikzpicture}[every node/.style={font=\footnotesize}]

\begin{scope}[xshift=0.0\picwidth,,yshift=+0.0\picheight]
\input{q_factor.tex}
\end{scope}
\setlength{\picwidth}{0.45\textwidth}
\setlength{\picheight}{0.45\textwidth}
\begin{scope}[xshift=-0.0\picwidth,,yshift=-.55\picheight]
%
%

\begin{axis}[%
width=0.4\picwidth,
height=0.4\picheight,
at={(0\picwidth,0\picheight)},
scale only axis,
axis on top,
xmin=-5.05050505050505,
xmax=5.05050505050505,
xlabel={$(b)$ $16-$QAM of $(C\boxplus D1)$},
xticklabels={},
xmajorgrids,
y dir=reverse,
ymin=-5.05050505050505,
ymax=5.05050505050505,
yticklabels={},
ymajorgrids,
axis background/.style={fill=white},
]
\addplot [forget plot] graphics [xmin=-5.05050505050505,xmax=5.05050505050505,ymin=-5.05050505050505,ymax=5.05050505050505] {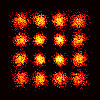};
\end{axis}
\end{scope}
\begin{scope}[xshift=0.48\picwidth,yshift=-.55\picheight]
%
%

\begin{axis}[%
width=0.4\picwidth,
height=0.4\picheight,
at={(0\picwidth,0\picheight)},
scale only axis,
axis on top,
xmin=-5.05050505050505,
xmax=5.05050505050505,
xlabel={$(c)$ $16$-QAM of $(C\boxplus D2)$},
xticklabels={},
xmajorgrids,
y dir=reverse,
ymin=-5.05050505050505,
ymax=5.05050505050505,
yticklabels={},
ymajorgrids,
axis background/.style={fill=white},
]
\addplot [forget plot] graphics [xmin=-5.05050505050505,xmax=5.05050505050505,ymin=-5.05050505050505,ymax=5.05050505050505] {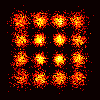};
\end{axis}
\end{scope}
\end{tikzpicture}
\vspace*{-10pt}
  \caption{Continuous spectrum of received pulses: $(a)$ Q-factor vs. $P$. $16-$ QAM constellation of received 64 subchannels for $(b)$ $(C\boxplus D1)$ and $(c)$ $(C\boxplus D2)$\label{fig:con_spec}}
\end{figure}
The transmission performance of $1460$ km link is shown in Fig.~\ref{fig:con_spec} 
in terms of the Q-factor directly derived from the bit-error-rate (BER), 
where more than 100 errors were obtained for each calculated BER value.
In all curves, the launch power is tuned by changing only 
the power of continuous spectrum, $P_{\rm c}$.

In comparison to conventional $64\times0.5$GBaud $16$-QAM OFDM, 
the NFDM (OFDM in continuous spectrum) shows about 0.7 dB gain in Q-factor
with a larger nonlinear threshold. Multiplexing with discrete spectrum, 
the nonlinear threshold (of continuous part) is increased by about 2 dB for 
$(C\boxplus D1)$ and by about 3 dB for $(C\boxplus D2)$. These values are consistent 
with the additional $P_{\rm d}$ of $(D1)$ and $(D2)$. 
Almost no decrease in Q-factor for  $(C\boxplus D1)$ indicates 
a low interference between the continuous and discrete spectrum in presence of
noise and other perturbations from ideal fiber model.
The performance degradation in $(C\boxplus D2)$ is mainly because of  large peak-to-average-power-ratio and limited 
resolution of ADC in receiver. The solitons keep their large
amplitudes but
the dispersive part of the pulse (continuous spectrum) 
spreads out into the guard bands and its effective amplitude decreases. Thus,
the dispersive part is captured with lower resolution than the case of without solitonic part.
\vspace{4pt}
\\
\emph{\textit{B. Discrete Spectrum:}}
The discrete spectrum of each symbol $\{(\hat{\lambda}_1,\hat{q}_d(\hat{\lambda}_1)),(\hat{\lambda}_2,\hat{q}_d(\hat{\lambda}_2))\}$ is computed via Forward-Backward NFT algorithm~\cite{aref2016control}. After eigenvalues detection, the encoded phase is detected by
the single-tap phase-shift removal operation performed according to designed eigenvalues~\cite{aref2015experimental}, and
applying the phase noise estimation derived for continuous part.

\begin{figure}[tb!]
\setlength{\picwidth}{0.5\textwidth}
\setlength{\picheight}{0.5\textwidth}
  \centering
\input{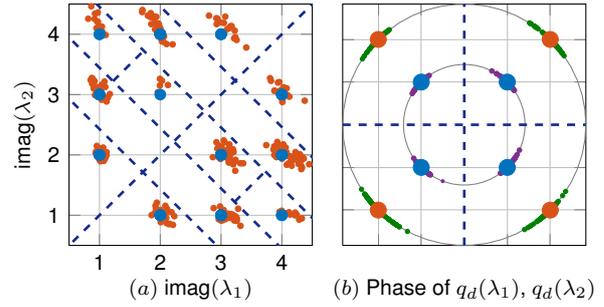}
\vspace*{-10pt}
  \caption{Discrete spectrum of received pulses for $(C\boxplus D2)$: $(a)$ imaginary part of received eigenvalues  and $(b)$ $\sphericalangle q_d(\lambda_k)$}
\label{fig:dis_spec}
\vspace*{-.1cm}
\end{figure}

For $(C\boxplus D2)$, the discrete spectrum of 300 received symbols is shown in Fig.~\ref{fig:dis_spec} transmitted at 
the nonlinear threshold. The two independent $q_d(\lambda_k)$ are depicted in the same plot.
The dots represent the received spectrum,
and the circles are the sent constellation points. The normalized error of eigenvalues
$$
(e_1,e_2)=\textstyle\frac{1}{\text{imag}(\lambda_1+\lambda_2)}(\hat{\lambda}_1-\lambda_1,\hat{\lambda}_2-\lambda_2)
$$ 
are  correlated $corr(e_1,e_2)\approx -0.499 -0.03j$ implying
$$
\hat{\lambda}_1 + \hat{\lambda}_2 \approx (\lambda_1 + \lambda_2) (1 + \epsilon), 
$$  
where $\epsilon$ is a small random variable with $\mathbb{E}(|\epsilon|^2)< 10^{-3}$. It suggests $(i)$ the decision regions as shown by dashed line in Fig.~\ref{fig:dis_spec} (a), and more importantly, $(ii)$ there is a small power exchange between discrete and continuous spectrum. 
By fitting a Gaussian mixture model, the BER is estimated resulting Q-factor beyond $14$ dB. A higher Q-factor is also estimated for both phases of $q_d(\hat{\lambda}_1)$ and $q_d(\hat{\lambda}_2)$ shown in Fig.~\ref{fig:dis_spec} (b). Thus,
the achievable rate is very close to the maximum $4+\log_2(12)$ information bits (error-free channel).

Similar performance is observed for other launch powers and also for the case $(C\boxplus D1)$ indicating an almost error-free transmission.

\section{Conclusions}
For the first time, the modulation of both continuous spectrum and discrete spectrum was demonstrated  in experiment. We observed that:
$(i)$ The continuous and discrete spectrum
are ``almost'' uncorrelated.
$(ii)$
The nonlinear threshold of continuous spectrum increases by about the power of discrete spectrum.
$(iii)$ The large PAPR of the discrete part  increases the quantization error of ADC which degrades the performance.
$(iv)$ Encoding more bits on discrete part seems quite feasible increasing the total SE.


\bibliographystyle{abbrv}


\vspace{-4mm}

\end{document}